\documentclass[acmtog]{acmart}
\usepackage{booktabs} 
\usepackage{bm} 
\usepackage{subcaption} 

\newcommand{\rev}[1]{#1}

\let\oldhat\hat
\renewcommand{\vec}[1]{\mathbf{#1}}
\renewcommand{\hat}[1]{\oldhat{\mathbf{#1}}}

\setcopyright{acmcopyright}
\acmJournal{TOG}
\acmYear{2018}\acmVolume{37}\acmNumber{6}\acmArticle{197}\acmMonth{11} \acmDOI{10.1145/3272127.3275031}

\acmSubmissionID{206}

\begin{document}

\title[A System for Acquiring, Processing, and Rendering Panoramic Light Field Stills for Virtual Reality]{A System for Acquiring, Processing, and Rendering Panoramic Light Field Stills for Virtual Reality}

\author{Ryan S. Overbeck}
\affiliation{%
  \institution{Google Inc.}
}

\author{Daniel Erickson}
\affiliation{%
  \institution{Google Inc.}
}

\author{Daniel Evangelakos}
\affiliation{%
  \institution{Google Inc.}
}

\author{Matt Pharr}
\affiliation{%
  \institution{Google Inc.}
}

\author{Paul Debevec}
\affiliation{%
  \institution{Google Inc.}
}



\begin{abstract}
We present a system for acquiring, processing, and rendering panoramic light field still photography for display in Virtual Reality (VR).  We acquire spherical light field datasets with two novel light field camera rigs designed for portable and efficient light field acquisition.  We introduce a novel real-time light field reconstruction algorithm that uses a per-view geometry and a disk-based blending field.  We also demonstrate how to use a light field prefiltering operation to project from a high-quality offline reconstruction model into our real-time model while suppressing artifacts.  We introduce a practical approach for compressing light fields by modifying the VP9 video codec to provide high quality compression with real-time, random access decompression.  

We combine these components into a complete light field system offering convenient acquisition, compact file size, and high-quality rendering while generating stereo views at 90Hz on commodity VR hardware.  Using our system, we built a freely available light field experience application \rev{called \emph{Welcome to Light Fields}} featuring a library of panoramic light field stills for consumer VR which has been downloaded over 15,000 times.

\end{abstract}

\begin{teaserfigure}
\centering
\includegraphics{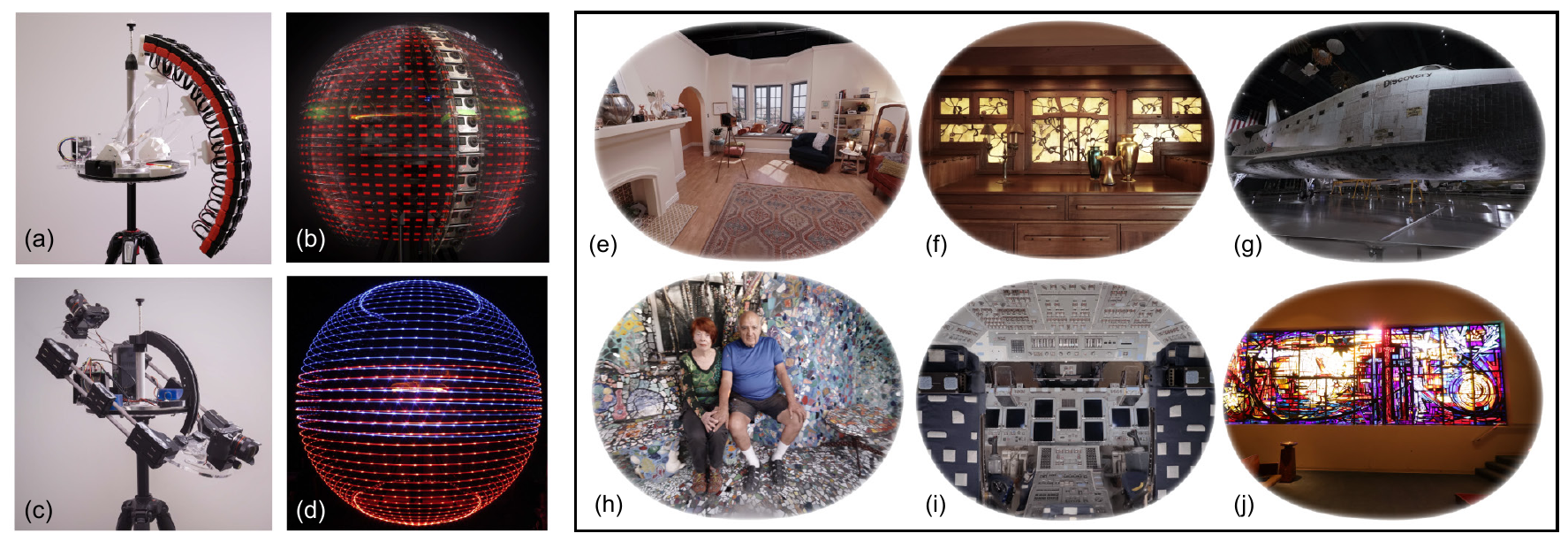}
\caption{Acquiring and rendering panoramic light field still images.  (a,c) Our two rotating light field camera rigs: the 16$\times$GoPro rotating array of action sports cameras (a), and a multi-rotation 2$\times$DSLR system with two mirrorless digital cameras (c).  (b,d) Long exposure photographs of the rigs operating with LED traces indicating the positions of the photographs.  (e-j) Novel views of several scenes rendered by our light field renderer.  The scenes are: (e) \emph{Living Room}, (f) \emph{Gamble House}, (g) \emph{Discovery Exterior}, (h) \emph{Cheri \& Gonzalo}, (i) \emph{Discovery Flight Deck}, and (j) \emph{St. Stephens}.}
\label{fig:teaser}
\end{teaserfigure}

%
%



\maketitle

\section{Introduction}
\label{sec:intro}

\rev{The advent of consumer VR headsets such as HTC Vive, Oculus Rift, and Windows Mixed Reality has delivered high-quality VR technology to millions of new users.  Notably, these headsets support positional tracking, which allows VR experiences to adjust the perspective shown by the headset according to how the user moves their head, creating experiences which are more immersive, more natural, and more comfortable.  To date, however, most photographically-acquired content such as 360 Video and Omnidirectional Stereo video can only be displayed from a fixed position in space and thus fails to benefit from positional tracking.}

\rev{Panoramic light fields provide a solution for acquiring and displaying real-world scenes in VR in a way that a user can not only look in every direction, but can move their head around in the scene as well.  As a result, the user receives a more comfortable and immersive experience, and even gets a better sense of the materials within the scene by observing properly reproduced view-dependent surface reflections.}


\rev{Light Fields also offer benefits over traditionally-modeled 3D scenes, which typically require extensive effort for artists to build detailed models, author complex surface shaders, and compose and simulate complex lighting.}  Rendering times for modern global illumination algorithms take hours per frame, as render farms perform trillions of ray queries, texture lookups, and
shader evaluations.  For scenes available to be photographed, light fields vastly reduce this effort,
achieving photorealism by simply interpolating and extrapolating pixel data from images which are already photorealistic.  Moreover, their simplicity suggests rendering solutions that can be measured in milliseconds, independent of scene complexity, achieving the rates needed for compelling virtual reality.


Although light fields provide many benefits, they have yet to appear in many commercial applications despite being introduced to computer graphics over 20 years ago~\cite{levoy1996light,gortler1996lumigraph}.  We believe there are four primary reasons for this:

\begin{enumerate}
\item \emph{Use-case:}
A compelling product use-case must exist in order for both content creators and distribution platforms to support the production of light field imagery.
\item \emph{Acquisition complexity:}
Creating high quality light fields requires acquiring thousands of images from a dense set of viewpoints.  A light field camera must capture all of these images in a reasonable amount of time, while also being portable and reliable.
\item \emph{Size:}
Light fields comprising thousands of images are unreasonably cumbersome for storage and transmission in uncompressed form.

\item \emph{Quality:} 
In order to maintain immersion, light field rendering must be free of distracting artifacts.  
\end{enumerate}


Speaking to (1), the advent of high-quality consumer VR technology with positional head tracking provides us with a new and compelling opportunity for viewing light fields, able to stimulate and satisfy our important sense of motion parallax.  Moving one's head in a scene not only reveals its depth in a compelling way, but it reveals information about the materials and lighting in a scene as we watch reflections shift over and play off of the scene's surfaces.
In this way, light fields are an ideal photo and video format for VR, as an HMD allows one to step inside of a lightfield photo or video and be teleported to far away and exotic places.  To maximize the feeling of presence, our work focuses on panoramic light field still images.  

The medium of VR presents another challenge:

\begin{enumerate}
\setcounter{enumi}{4}
\item \emph{Speed:}
True immersion in VR requires rendering at exceedingly high framerates.  While 30Hz or 60Hz may have been enough for
traditional real-time applications, current desktop VR platforms require applications to render at 90Hz to maximize user comfort.
\end{enumerate}

Our system overcomes the technical challenges (2)-(5) above with several novel advances in light field acquisition, processing, and rendering.  We acquire our light fields with one of our two novel light field camera rigs (see Figure~\ref{fig:teaser}(a,c)).  These rigs capture thousands of images on the surface of the sphere (see Figure~\ref{fig:teaser}(b,d)) and were designed with efficiency and portability in mind.  We then feed the images into our cloud-based processing pipeline.  The pipeline calibrates the images to determine the real-world locations of the cameras and extracts depth maps to be used by our renderer.  Also during processing, we perform a novel application of light field prefiltering that improves final reconstruction quality by smoothing away distracting ghosting artifacts.  The pipeline finishes by compressing the data using a practical codec for light field stills.  Based on VP9~\cite{VP9}, a mainstream open source video codec, our codec achieves light field compression rates competitive with video while also retaining real-time random access to individual image tiles.  We render these compressed light fields using our novel real-time light field reconstruction algorithm.  This algorithm uses a per-view geometry to perform depth correction and blends the multiple views using a disk-based blending field.  We achieve high quality real-time rendering (see Figure~\ref{fig:teaser}(c-j)) with relatively sparse light field data sets.

The result is a complete system for acquiring, processing, and rendering lightfield still imagery that produces high quality reconstructed views and runs at 90Hz on entry-level VR-enabled machines.  Moreover, the datasets compress down to 50-200MB with few noticeable compression artifacts, so that a user can download a light field still in a matter of
seconds on a common 10-20 Mbit/s internet connection.  \rev{We have launched this system to the public as \emph{Welcome to Light Fields}, a freely downloadable application on the Steam$\circledR$ store (https://store.steampowered.com/).  \emph{Welcome to Light Fields} allows people to virtually explore places, including the scenes in this paper, in a deeply immersive way.}

After a review of related work in Section~\ref{sec:related}, the rest of this paper focuses on these primary contributions:
\begin{itemize}
    \item two novel light field acquisition devices that can acquire dense panoramic light field stills and are portable and efficient (Section~\ref{sec:acquisition}),
    \item a novel real-time light field reconstruction algorithm that produces higher quality results with sparser light field images than previous real-time approaches (Section~\ref{sec:reconstruction}),
    \item a novel application of light field prefiltering to reduce many light field rendering artifacts (Section~\ref{sec:prefilter}),
    \item a practical light field codec implemented using VP9 that provides aggressive compression while also retaining real-time random access to individual image tiles (Section~\ref{sec:compression}), and
    \item taken altogether, a system that provides a complete solution to acquiring, processing, and rendering panoramic light field stills for VR (Section~\ref{sec:rendering}).
\end{itemize}


\section{Related Work}
\label{sec:related}
There has been significant research on light fields which is well covered by surveys~\cite{wu2017light,zhang2004survey}.  In this section, we focus on the works that are most relevant to understanding our system.

\paragraph{Panoramic light fields for VR}
A few recent projects \cite{Debevec:2015:SLF,milliron2017hallelujah} have pointed toward the power of panoramic light fields in VR.  Our application goals are similar, but in this work we aim to document a complete system for light field still acquisition, processing, compression, and rendering which can produce downloadable experiences on even entry-level consumer VR hardware.

\rev{There are several other image based rendering (IBR) approaches that aim for a simpler representation than light fields by reducing the dimensionality of the data, but in doing so restrict the space of the viewer's motion.  Omnidirectional stereo~\cite{anderson2016jump,konrad2017spinvr} provides stereo views for all directions from a single view point.  Concentric mosaics~\cite{shum1999rendering} allow movement within a circle on 2D plane.  Our panoramic light fields provide the full 6 degrees of freedom (6DOF) of movement (3 for rotation and 3 for translation) in a limited viewing volume.  This leads to a more comfortable and fully immersive viewing experience in VR.}

\rev{
Light fields have also proven useful to render effects of refocus~\cite{ng2005light} and visual accomodation~\cite{lanman2013near}.  Birklbauer and Bimber~\shortcite{birklbauer2014panorama} extended this work to panoramic light fields.  However, these approaches require extremely dense image sets acquired using microlens arrays, and allow viewpoint changes of only a few millimeters.  We record larger volume light fields for stereoscopic 6DOF VR by moving traditional cameras to different positions as in Levoy and Hanrahan~\shortcite{levoy1996light}, and we use depth-based view interpolation to overcome the resulting aliasing of the light field from the greater viewpoint spacing.}  As future work, it could be of interest to display such datasets on neareye light field displays to render effects of visual accommodation.

We render our high quality panoramic light fields with a novel disk-based light field reconstruction basis in Section~\ref{sec:reconstruction}.
Similar to the work of Davis et al.~\shortcite{davis2012unstructured} and Buehler et al.~\shortcite{buehler2001unstructured}, our basis supports loosely unstructured light fields: we expect the light field images to lie on a 2D manifold and achieve best performance when the images are evenly distributed.  \rev{We use a disk-based basis in order to render with a per-view geometry for high quality reconstruction.  Davis et al.~\shortcite{davis2012unstructured} proposed a disk-based method specifically for wide-aperture rendering, but their approach doesn't immediately allow rendering with per-view geometry as ours does.  Our approach also allows wide-aperture rendering while offering more flexibility in aperture shape.}

Our system leverages depth-based view interpolation to render rays of light field, even when the rays do not pass through the original camera centers.  There are offline view interpolation algorithms that can achieve high quality rendering of light field data\rev{~\cite{penner2017,zitnick2004high,shi2014light}}.  Of particular interest are recent efforts to use machine learning e.g. ~\cite{kalantari2016learning}.  Although these approaches are currently limited to offline usage, we expect them to become more relevant to real-time applications such as ours in the near future.


\paragraph{Light field acquisition} Most light fields have been acquired with motorized camera gantries (e.g. \cite{levoy1996light}), hand-held cameras (e.g. \rev{\cite{gortler1996lumigraph,davis2012unstructured,birklbauer2015active}}), or imaging through lenslet arrays (\cite{ng2005light}).  Lenslet arrays typically capture too small a viewing volume for VR, and motorized gantries are usually slow or optimized for planar or inward-pointing data.  In our work, we present two mechanical gantry camera rigs which capture outward-looking spherical light field datasets in a matter of minutes or as little as 30 seconds, with greater repeatability than hand-held solutions.

\paragraph{Geometry for improved light field reconstruction}
It has been demonstrated that image-based rendering (IBR) techniques, including light fields, can be improved by projecting sources images onto a geometric model of the scene~\cite{lin2004geometric,chai2000plenoptic}.  Most previous work uses a single global scene model~\cite{levoy1996light,gortler1996lumigraph,buehler2001unstructured,davis2012unstructured,shade1998layered,debevec1998efficient,wood2000surface,chen2002light,sloan1997time}.  More recently, it has been shown that using multi-view or \emph{per-view} geometries can provide better results~\cite{zitnick2004high,hedman2016scalable,penner2017} by optimizing each geometry to best reconstruct a local set of views.  However, most of these approaches target offline view synthesis~\cite{zitnick2004high,penner2017}.  \rev{There are several real-time approaches that target free-viewpoint IBR by using multi-view or per-view geometries~\cite{hedman2016scalable,eisemann2008floating,chaurasia2013depth}}.  Relative to their work, we focus on denser light field datasets, sacrificing some range of motion in order to more faithfully reproduce reflections and other view-dependant effects when viewed from within the light field volume.


\paragraph{Light field compression}
Viola et al.~\shortcite{viola2017comparison} provide a thorough review of many approaches to light field compression.  Most approaches, including ours, build on the insight that the disparity between images in a light field is analogous to the motion between neighboring frames in a video~\cite{lukacs1986predictive}, and so similar encoding tools may be used.  Most previous approaches require the full light field to be decoded before any of it can be rendered.  However, for immersive stereoscopic viewing in VR, a relatively small portion of the light field is visible at any time.  Based on this insight, we decided to focus on compression techniques that allow for random access to light field tiles and on-demand decoding\rev{~\cite{levoy1996light,zhang2000compression,magnor2000data}}.

The basis of our compression algorithm, as described in Subsection~\ref{sec:lf-compress}, is similar to the work of Zhang and Li~\shortcite{zhang2000compression}.  The primary impact of our work lies in implementing this approach in an existing video codec, VP9~\cite{VP9}, and demonstrating its use in a high quality light field rendering algorithm.  

\section{Acquisition}
\label{sec:acquisition}


We built two camera rig systems designed to record the rays of light incident upon a spherical light field volume.

\paragraph{16$\times$GoPro rig.} Our first rig, prioritizing speed, is built from a modified GoPro Odyssey omnidirectional stereo (ODS) rig \cite{anderson2016jump} to create a vertical arc of 16 GoPro Hero4 cameras (Figure~\ref{fig:teaser}(a)).  The camera arc is placed on a rotating platform which has a custom-programmed PIC board to drive a stepper motor to spin the array in a full circle in either 30, 60, or 120 seconds.  For the longer settings, the rig can support both continuous rotation, or a "start-stop" mode which brings the motion of the arc to a halt at 72 positions around the circle.  The fastest 30 second rotation time is well suited for brightly lit conditions when motion blur is not a problem, and for capturing people.  The slower rotation times and start-stop mode were used to avoid motion blur when the cameras were likely to choose long exposure times.

By extracting evenly-spaced video frames from the 16 video files ($2704 \times 2028$ pixels, 30fps), we obtain a dataset of 16 rows of 72 pictures around, or 6 Gigapixels.  (We drop some of the clumped-together images near the poles to even out the distribution.)  For closer subjects, we can double the number of rows to 32 by rotating the camera array twice.  In this mode, the top of the camera arc is attached to a string wound around a vertical rod, and the unwinding of the string allows the arm to pivot down along its arc by half the camera spacing each rotation (Figure~\ref{fig:teaser}(b)).

\paragraph{2$\times$DSLR rig.} Our second rig, prioritizing image quality, spins a pivoting platform with two Sony a6500 mirrorless cameras (Figure~\ref{fig:teaser}(c)).  Their 8mm Rokinon fisheye lenses point outward from opposite sides of an 80cm long platform, and one side is additionally weighted.  A similar drop-string mechanism as before allows the cameras to pivot lower/higher approx. 3cm each time the platform rotates, sending one camera in an upward spherical spiral and the other downward (Figure~\ref{fig:teaser}(d)).  The control board triggers the cameras to take a photo 72 times each rotation, producing spherical photo sets with 18 rows and 72 columns.  The drop-string allows the cameras to cover the whole sphere of incident viewing directions with a single motor mechanism, reducing the cost and complexity of the system.

The mirrorless cameras support "silent shooting" with electronic (rather than physical) shuttering.  This allows fast shooting of bracketed exposures for high dynamic range imaging.  When the rig is run in start-stop mode, HDR images of -2, 0, and +2 stops can be recorded at 1.5 seconds between positions, yielding 7,776 photos total in 33 minutes.  The rig can also be run continuously, yielding 2,592 single-shot photos in nine minutes.  When running continuously, the shutter speed on each camera must be set to 1/250th second exposure or less in order to produce sufficiently sharp\footnote{$1/250^{th}$ sec shutter rotating at 2 rpm turns the cameras $1/7500^{th}$ of a rotation during exposure, allowing light fields of $7500/360 \approx 20$ pixels per degree at the equator.} images at the fastest rotation setting of 30 seconds.

The diameter of the capture volume was chosen to provide enough room for acceptable seated head movement while keeping the rigs relatively light and maneuverable.  The GoPro-based rig sweeps the camera centers along the surface of a sphere 70cm in diameter, which, given the GoPro's 110$^\circ$ horizontal field of view, yields a light field viewing volume 60cm in diameter, and slightly smaller in height (the mirrorless camera rig's volume is slightly greater at 70cm).  We found that this was a sufficient volume to provide satisfying parallax for a seated VR experience, and it also allowed us to fit the rig into tighter spaces, such as the flight deck and airlock of the space shuttle.

\section{Light field Reconstruction with Disk-based Blending Field and Per-view Geometry}
\label{sec:reconstruction}

\begin{figure}
    \centering
    \includegraphics{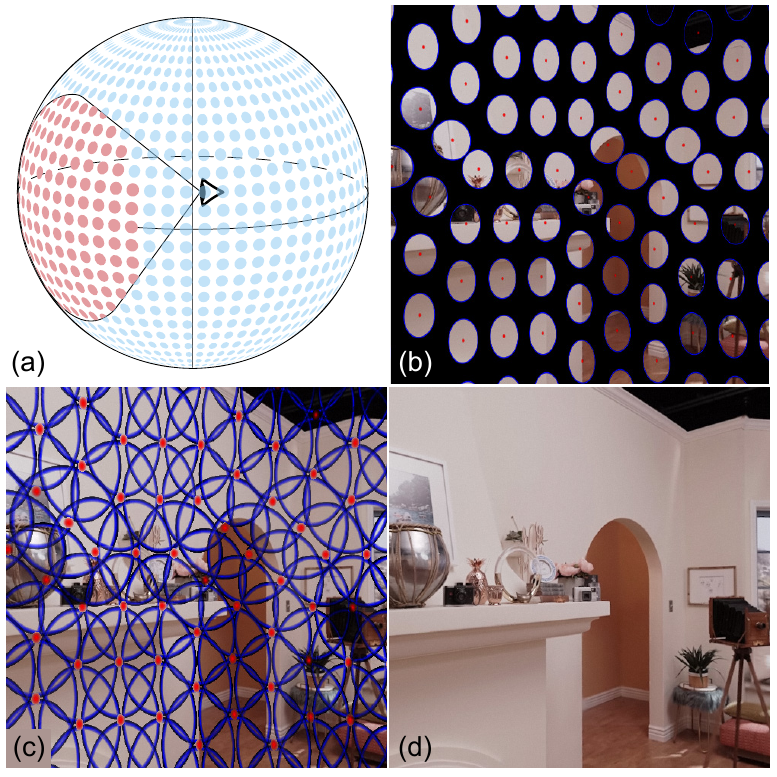}
    \caption{We use a disk-based blending field for light field reconstruction.  (a) The disks are positioned on the capture surface.  (b) To generate a novel view, the disks are projected onto the novel view's image providing windows onto each light field image.  (c) The disks are sized large enough to overlap into a seamless view of the scene (d).}
    \label{fig:disks-shrink}
\end{figure}

To render our light fields, we project the images onto a per-view geometry and blend the results using a disk-based reconstruction basis as seen in Figure~\ref{fig:disks-shrink}.  Each light field image is represented by a disk (Figure~\ref{fig:disks-shrink}(a)), and each disk acts as a window onto the image (Figure~\ref{fig:disks-shrink}(b)).  The size of the disks determines the amount of blending between neighboring views, and we choose large enough disks to blend together a seamless view of the light field (Figure~\ref{fig:disks-shrink}(c, d)).

Figure~\ref{fig:disk-proj} illustrates how the contents of each disk are rendered using a per-view geometry proxy.  Each light field image is projection mapped onto a distinct triangle mesh derived from a multiview stereo depth map.  The resulting textured mesh is masked and weighted by the disk and projected onto the screen for the novel view.  
These separate textured meshes are blended together as in Figure~\ref{fig:disks-shrink}.

As highlighted in Section~\ref{sec:vs-camera-mesh}, the primary benefit of our disk-based representation is that it makes it efficient to render the view from each light field image as a separate textured triangle mesh and blend the views together all on the GPU.  Rendering with these per-image geometry proxies allows for better reconstruction than using a single global geometry proxy.

In the following subsections, we describe the details of our algorithm. 
We start with a higher level mathematical description in Subsection~\ref{sec:render-disks} before describing the GPU implementation in Subsection~\ref{sec:gpu}.  We emphasize how our approach improves upon traditional camera mesh based light field rendering used by most previous methods in Subsection~\ref{sec:vs-camera-mesh}.  Finally, we detail how we generate the size and shapes of the disks in Subsection~\ref{sec:disk-shape} and the meshes for each image in Subsection~\ref{sec:mesh-gen}. 




\subsection{Rendering with Disks}
\label{sec:render-disks}

\begin{figure}
    \centering
    \includegraphics{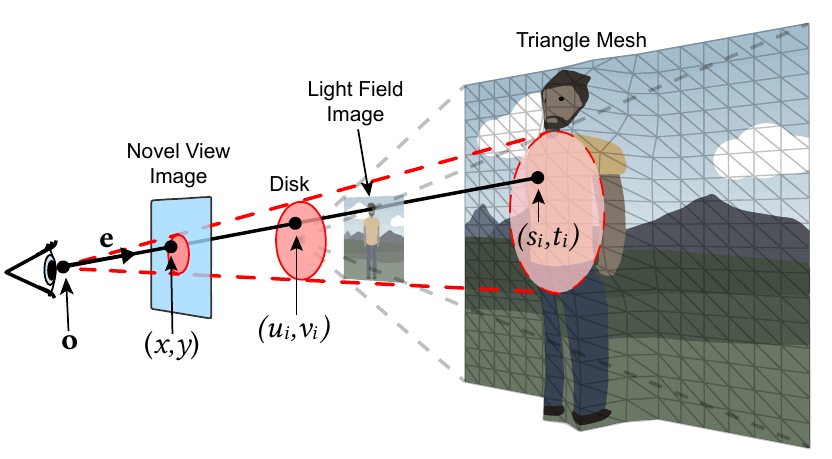}
    \caption{To compute $I_i(s_i,t_i)$ in Equation~\ref{eq:lfsplat}, each light field image is projection mapped onto a mesh which is masked and weighted by the disk and projected onto the image for the novel view.}
    \label{fig:disk-proj}
\end{figure}

Figure~\ref{fig:disk-proj} illustrates our disk-based rendering algorithm.
We compute the image for a novel view of the light field, $\Phi(x,y)$, by performing ray look ups into the light field data:
\begin{equation}
    \Phi(x,y)=L(\vec{o}, \vec{e}).
    \label{eq:image-ray}
\end{equation}
$L(\vec{o}, \vec{e})$ is a light field ray query with ray origin $\vec{o}$ and direction $\vec{e}$.  In the context of Equation~\ref{eq:image-ray}, $\vec{o}$ is the center of projection for the novel image, and $\vec{e}$ is the vector from $\vec{o}$ through the pixel at $(x,y)$  (see Figure~\ref{fig:disk-proj}). 
Although light fields are generally parameterized as four-dimensional, it is more convenient for our purposes to use this 6D ray query.

We use a disk-based representation to answer this ray query:
\begin{equation}
L(\vec{o}, \vec{e}) = \frac{\sum_{i} I_i(s_i,t_i) \rho_i(u_i,v_i)}{\sum_i \rho_i(u_i,v_i)},
\label{eq:lfsplat}
\end{equation}
where $I_i$ draws a continuous sample from the light field image $i$, and $\rho_i$ maps a position on the disk to a weight.  $(s_i, t_i)$ and $(u_i, v_i)$ are both functions of the ray, $(\vec{o}, \vec{e})$, and, as shown in Figure~\ref{fig:disk-proj}, they are at the intersection point between the ray and the triangle mesh and between the ray and the disk respectively.  $(s_i, t_i)$ are in the local coordinates of the light field image, and $(u_i, v_i)$ are in the local coordinates of the disk.


$\rho_i$ maps a position on the disk, $(u_i, v_i)$, to a weight in a kernel.  
The kernel can be any commonly used reconstruction kernel that is at its maximum at the center of the disk and goes to zero at the edge.  We use a simple linear falloff from 1 at the disk center, but it is also possible to use a Gaussian, spline, Lanczos, or other kernel.  Each disk at $i$ is centered at the center of projection of each image at $i$ and oriented tangential to the spherical light field surface (see Figure~\ref{fig:disks-shrink}(a)).  The size and shape of each disk is computed to achieve an appropriate amount of overlap between neighboring disks.

Note that Equation~\ref{eq:lfsplat} derives from the formula for point splatting~\cite{gross2011point,zwicker2001surface}:
\begin{equation}
\Phi(x,y)=\frac{\sum_{i} c_i \rho_i(u_i,v_i)}{\sum_i \rho_i(u_i,v_i)},
\label{eq:splat}
\end{equation}
where instead of using the constant splat color, $c_i$, in Equation~\ref{eq:splat}, we sample the light field image, $I_i(s_i,t_i)$, in Equation~\ref{eq:lfsplat}.

\subsection{GPU Implementation}
\label{sec:gpu}
Equation~\ref{eq:lfsplat} is implemented as a 2-pass algorithm in OpenGL.  In the first pass, the numerator is accumulated in the RGB components of an accumulation framebuffer while the denominator is accumulated in the alpha component.  In the second pass, we perform the division by rendering the accumulation framebuffer as a screen aligned quad and dividing the RGB components by the alpha component.  Most of the work is done in the first pass, so we focus on that in the following.  

In the first pass, we render texture mapped triangle meshes in order to compute $I_i(s_i,t_i)$ in Equation~\ref{eq:lfsplat}.  The triangle meshes are sent to the GPU via the OpenGL vertex stream, and the vertices are expanded, transformed, and rendered in the shader program.  The light field image data is sampled from a texture cache, which is described below in Subsection~\ref{sec:tile-streaming}.  The texture coordinates at the vertices are passed along with the vertex stream.  

The disk blending function, $\rho_i(u_i,v_i)$, is also computed in the first pass shader program.  We compute the intersection between the ray through the fragment at $(x, y)$ and the disk for image $i$, and then sample the kernel at the intersection point.  The disk parameters are stored in a \emph{shader storage buffer object} (SSBO), which is loaded during program initialization before rendering.  

After the shader program finishes with a fragment, the results are accumulated to the framebuffer to compute the sums in Equation~\ref{eq:lfsplat}.  The sums are accumulated in a 16-bit floating point RGBA framebuffer using simple additive blending.  The accumulation framebuffer is passed as a texture to the second pass which performs the division as explained at the beginning of this subsection.

\subsubsection{Tile Streaming and Caching}
\label{sec:tile-streaming}
If implemented na\"\i vely, the algorithm as described above will render all of every textured mesh for all light field images every frame, requiring the entire light field to be loaded into GPU memory.  This would be prohibitively expensive for our typical light field resolution.  
To avoid this, we use a tile streaming and caching architecture similar to that used in previous work~\cite{birklbauer2013rendering,zhang2000compression}.  

First, during processing, we divide the light field images into smaller tiles.  We generally use 64$\times$64 tiles, but different tile sizes offer trade-offs between various computational costs in the system.  Both the light field imagery and the triangle meshes are divided into tiles.  

At render time, we perform CPU-based culling to remove any tiles that aren't visible through their disks.  For each visible tile, we load the image tile into a GPU tile cache if it's not already there, and we append the tile's mesh to the vertex array that is streamed to the GPU shader program described above in Subsection~\ref{sec:gpu}.  The tile cache is implemented in OpenGL as an \emph{array texture}, where each layer in the array is treated as a cache page.  We generally use a page size of 512$\times$512 or 1024$\times$1024 and concatenate tiles into those pages as needed.  The pages are evicted according to a simple FIFO scheme.  The address of the tile in the cache is streamed along with the mesh vertex data so that it can be used to fetch texture samples from the cache in the fragment shader.

Note that this tile streaming architecture has strong implications on how we compress the light field imagery, and we explain how we compress the light field while allowing for this per-tile random access in Section~\ref{sec:compression}.

\subsubsection{Rendering with Intra-Image Occlusion Testing}
\begin{figure}
    \centering
    \includegraphics{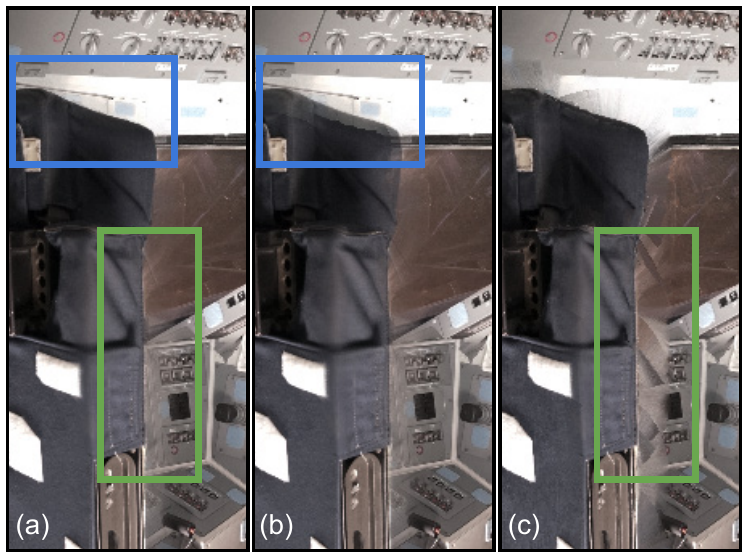}
    \caption{We use depth testing to mask out intra-image occlusions within each mesh and blend the results (a).  We found this to work better than turning off depth testing altogether (b), which results in artifacts where the background bleeds into the foreground, and depth testing between neighboring meshes (c), which causes stretched triangles at the edges to obstruct the view.  These close-ups are from the \emph{Discovery Flight Deck}.}
    \label{fig:occlusion}
\end{figure}

Each mesh is rendered using the GPU's depth test to mask out intra-image occlusions within the mesh.
As shown in Figure~\ref{fig:occlusion}, we found that intra-image occlusion testing (a) worked better than either testing for inter-image occlusions (c) or turning off the depth test altogether (b).  

To implement this, we could clear the depth buffer between processing each light field image, but this would be expensive.  Instead we scale the depth values in the shader program such that the depth output by the mesh from each light field image fill non-overlapping depth ranges
with the images drawn first using larger depth values.  
Thus meshes from images drawn later are never occluded by the previously drawn meshes.

\subsection{Comparison to Camera-Mesh Based Blending}
\label{sec:vs-camera-mesh}
\begin{figure}
    \includegraphics{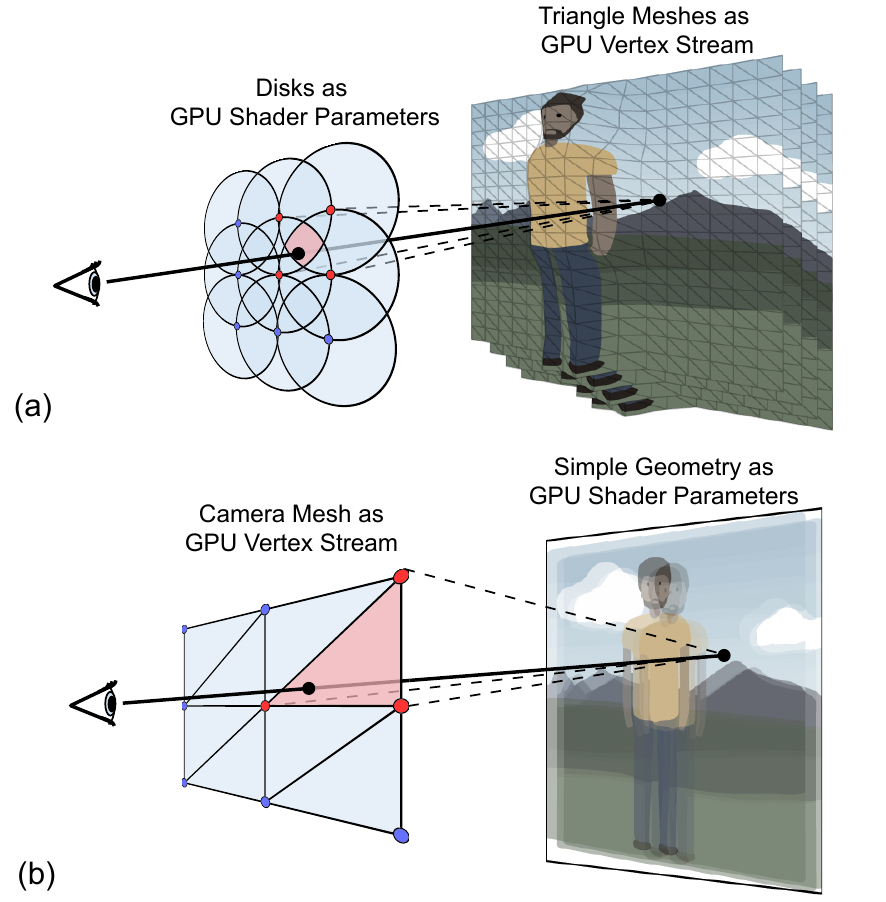}
    \caption{
    Our disks compared to the traditional camera mesh blending field. The red and blue dots are the centers of projection of the light field images.  The red dots are the images used to compute the given ray query.  The dashed lines are the rays used by each of the cameras to answer the ray query.
    (a) Our novel disk-based blending field encodes the disks compactly in shader parameters, freeing the triangle stream to express the scene geometry proxy. (b) The traditional approach uses the triangle stream to convey the camera mesh blending field and shader parameters to define the geometry proxy.}
    \label{fig:disk-render}
\end{figure}

Figure~\ref{fig:disk-render} compares our approach to a more classic method of light field rendering.
Figure~\ref{fig:disk-render}(b) shows the approach used by most previous real-time light field rendering algorithms~\cite{levoy1996light,gortler1996lumigraph,davis2012unstructured,buehler2001unstructured}. Instead of disks, they use a triangular mesh to define the reconstruction basis, where the mesh vertices are at the center of projection of the light field images, and the triangles designate triplets of images that get blended together via barycentric interpolation.  

In this classical approach, the triangles of the camera mesh are sent to the GPU via the triangle vertex stream.  This means that any scene geometry proxy must be communicated to the GPU shaders by some other means, usually as light-weight shader parameters.  Relative to our approach, this restricts the possible complexity expressed by the geometry proxy.  Often a single plane is used~\cite{levoy1996light}, or the geometry proxy is evaluated only at the vertices of the camera mesh~\cite{gortler1996lumigraph, buehler2001unstructured}.  Alternatively, one can render a single depthmap from a global geometry proxy in a separate pass and use it as a texture in a secondary pass that renders the camera mesh~\cite{davis2012unstructured}.  We tried to find a way to render a per-view geometry with the camera mesh blending field, but our solutions all required too many passes to be real-time.

Our disk-based algorithm in Figure~\ref{fig:disk-render}(a) transposes the use of the GPU triangle vertex stream and shader parameters:  the triangle stream carries the scene geometry proxy and the disk geometry is passed in shader parameters.  Intuitively, this makes rendering the geometry proxy more of a natural fit to the standard GPU pipeline.
Thus we are able to efficiently render with a per-view geometry using the algorithm described in Subection~\ref{sec:gpu}, which provides higher reconstruction quality than using a single global geometry.

Moreover, rendering using the camera mesh blending field makes it harder to implement the tile streaming and caching approach described in Subsection~\ref{sec:tile-streaming}.  Since it is not known until the fragment shader exactly which tiles to sample for each fragment, it is necessary to employ complicated shader logic that includes multiple dependant texture fetches which is expensive on current GPUs.  In our algorithm, each triangle rasterized by the GPU maps to exactly one tile, so we can pass the tile's cache location directly with the vertex data.


\subsection{Disk Shape}  
\label{sec:disk-shape}
\begin{figure}
\centering
\includegraphics{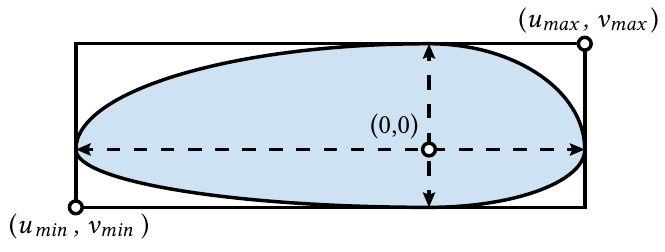}
\caption{We use asymmetric ovals for our disks.  Their shape is defined by the placement of the origin and an axis-aligned bounding box.}
\label{fig:disk-oval-rect}
\end{figure}
Our disks must be positioned and shaped such that they provide enough overlap for high quality reconstruction, while remaining compact enough to afford fast render performance.
We use ovals for our disks (see Figure~\ref{fig:disk-oval-rect}), though rectangles could also be used.  
Both shapes are asymmetric, and are defined by an origin point and a 2D axis-aligned bounding box.  \rev{The asymmetry allows for smaller disks and thus faster performance while maintaining sufficient overlap with neighbors.}  The asymmetric oval is the union of 4 elliptic sections, one for each quadrant around the origin.
The origin of each disk is at the center of projection of its corresponding light field image, the disk normal is aligned to the image's optical axis, and the $u,v$-dimensions are aligned with the $s,t$ image dimensions.  

To create an optimal light field reconstruction filter~\cite{levoy1996light}, the extents of each disk should reach the centers of the disks in its 1-ring according to a Delaunay triangluation of the disk centers.
We search for the furthest of the 1-ring neighbors in each of the $+u$, $-u$, $+v$, and $-v$ directions and use that to define the disk extents.  Disk size may be further tuned to trade between reconstruction quality and render speed.



\subsection{Mesh Generation}
\label{sec:mesh-gen}
As described above, we render a separate textured mesh for each light field image.  These meshes are pre-generated as part of our fully automated processing pipeline.  

We first compute a depth map for each light field image using an approach based on Barron et al.~\shortcite{Barron2015A}.  
Next the depth maps are tessellated with 2 triangles per pixel, and then the mesh is simplified to a reasonable size.  For mesh simplification, we use a modified version of the Lindstrom-Turk~\cite{lindstrom1998fast,lindstrom1999evaluation} implementation in CGAL~\cite{cgal:c-tsms-12-18a}.  As noted in Subsection~\ref{sec:tile-streaming}, we require a completely separate triangle mesh per light field image tile so that tiles can be streamed to the GPU independently.  Our modifications to CGAL's Lindstrom-Turk implementation guarantee that we get a separate mesh for each tile. 

We use a tunable parameter, \emph{geometry precision} ($gp$), to specify how aggressively to simplify the meshes.  Our algorithm simplifies the mesh until there are an average of $2\times gp$ triangles per tile.

\begin{figure*}
    \centering
    \includegraphics{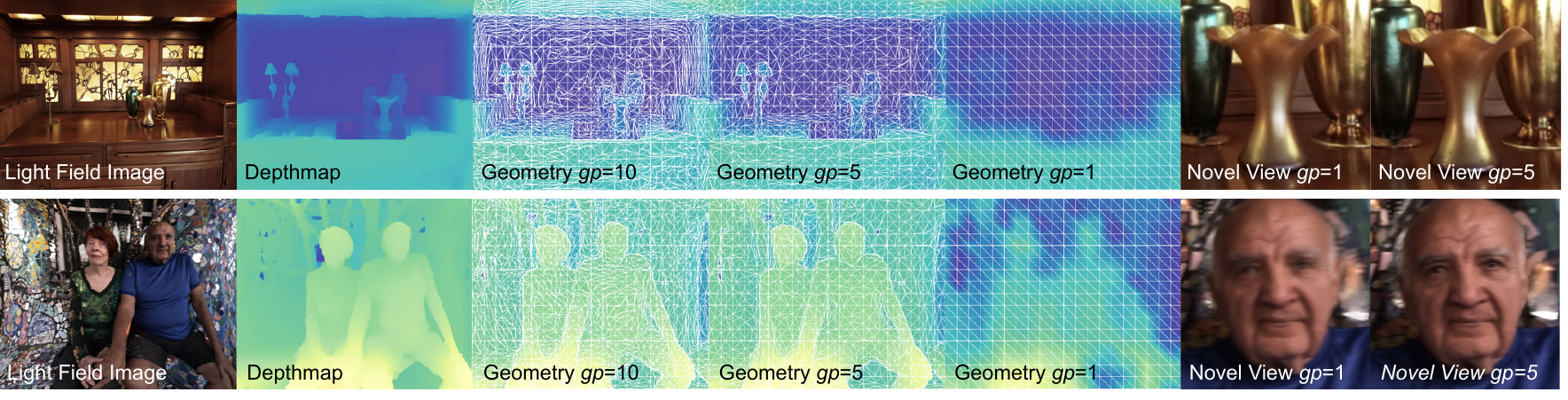}
    \caption{The impact of \emph{geometry precision} ($gp$) on reconstruction quality.  Shown from left to right is, a single image from the light field, the depthmap generated for that image, the simplified mesh generated from the depthmap at $gp=10$, $gp=5$, and $gp=1$, and the results of reconstructing a novel view at $gp=1$ and $gp=5$.  At $gp=1$, some ghosting is visible.  At $gp=5$, the results are sharp.  The reconstruction results for $gp=5$ and $gp=10$ are indistinguishable, so we leave out $gp=10$ results here.}
    \label{fig:geometry_res}
\end{figure*}

Figure~\ref{fig:geometry_res} shows some results at different $gp$ settings. At the lowest quality, $gp=1$, the meshes simplify down to 2 triangles per tile, and the reconstructed results show some visible ghosting artifacts.  At $gp=5$ there are an average of 10 triangles per tile, and the results are sharp.  Increasing $gp$ above 5 does not noticeably improve reconstruction results, so we set $gp=5$ by default, and all results in this paper use $gp=5$ unless explicitly stated otherwise.

When we store the meshes with the light field imagery, we quantize the $(s,t,depth)$ vertex coordinates to 8 bits per component and then use a variable length encoding.  Beyond that, we don't do any extra compression on the geometry data because it is still relatively small compared to the compressed light field image data.  The variable length encoding is decoded at load time, and the quantized vertex coordinates are streamed directly to the GPU, where they are expanded and projected to the screen in the shader program.

\section{Prefiltering to Improve Light Field Reconstruction}
\label{sec:prefilter}
Our disk-based light field reconstruction algorithm (Section~\ref{sec:reconstruction}) works well when the scene geometry is accurate, but can produce artifacts around depth edges where the geometry proxy deviates from the shape of the underlying scene.  And these artifacts can flicker from frame to frame as the user moves around.

We can significantly diminish these artifacts by applying a novel form of prefiltering to the light field imagery during processing.  The concept of light field prefiltering was introduced by Levoy and Hanrahan~\shortcite{levoy1996light} as a way of smoothing away frequencies that are not captured by a relatively sparse camera array.  To perform optical prefiltering during the acquisition process, one would need a very large camera aperture spanning the space between neighboring camera positions, which is difficult to achieve with very wide-angle lenses.  However, we can apply synthetic prefiltering to our imagery by modeling a large synthetic aperture camera.

We prefilter our light field imagery by applying a synthetic prefilter after the light field is acquired but before it is encoded for rendering.  To do so, we treat the original light field as a synthetic scene and acquire a new light field from it using cameras with apertures large enough to band-limit the 4D light field signal.
Since this is done offline, we can use a higher quality representation of the light field as the original light field representation and use the prefilter to project this higher quality light field onto our real-time light field model.

Of course, since this prefilter is applied after acquisition, we can't remove all aliased frequencies from the data.  However, by projecting from a higher quality offline model of the light field, we can remove spurious frequencies introduced by our simplified real-time model.  We have also found that prefiltering helps reduce the impact of other distracting artifacts as well.  

\subsection{Prefilter Implemention}
\label{sec:prefilter-impl}
\begin{figure}
    \centering
    \includegraphics{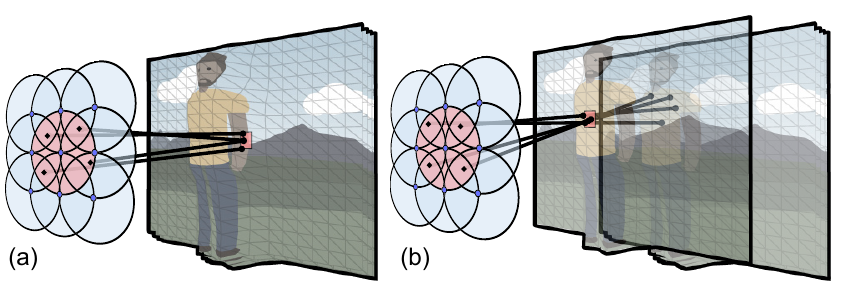}
    \caption{Our prefilter is a 4D integral between the light field disk (red oval) and the light field image pixel (red square).  (a) When the geometry proxies are close, the rays converge and the prefilter results are sharp.  (b) When the geometry proxies differ, the rays diverge, and the results are blurred.}
    \label{fig:prefilter}
\end{figure}

To implement the prefilter, we loop over all of the light field image pixels and compute a Monte Carlo approximation of the 4D integral over the 2D surface of the pixel and the 2D synthetic aperture for the camera (see Figure~\ref{fig:prefilter}).  For the synthetic aperture, we simply use the disks from our disk-based reconstruction of Section~\ref{sec:reconstruction} as these are already sized and shaped appropriately.  

The integral we aim to approximate to compute the prefiltered value at each light field image pixel, $I'_i(s',t')$, is given by:
\begin{equation}
    I'_i(s',t') = \iint_{D_i} \iint_{P_{i,s',t'}} \rho_i(u, v)\, \sigma(s, t)\, \bar{L}(\vec{o}, \vec{e}) \,
    ds\,dt\,\,du\,dv\,,
    \label{eq:prefilter-integral}
\end{equation}
where $\rho_i(u, v)$ is the disk shaped kernel as defined in Section~\ref{sec:render-disks}.  $\bar{L}(\vec{o}, \vec{e})$ is the light field ray query using a high quality offline light field reconstruction model, and $\sigma(s, t)$ is the pixel reconstruction basis function.  
The integrals are taken over the area of the disk, $D_i$, and the area of the pixel reconstruction basis function, $P_{i,s',t'}$.  The ray origin $\vec{o}$ is the 3D point on the disk at $(u,v)$ and the direction $\vec{e}$ points toward the point in the image at $(s,t)$.  The 3D location of the point at $(s,t)$ is the projection mapped location of that image point on the triangle mesh for light field image $i$.

For the pixel reconstruction basis function, $\sigma(s, t)$, we found that the Lanczos-2 kernel helped avoid both ringing and over-blurring.  We found that a simple box filter also works, but adds additional blur.

The Monte Carlo approximation of Equation~\ref{eq:prefilter-integral} using $J$ samples is given by:
\begin{equation}
    \tilde{I}'_i(s', t') = \frac{\sum_{j<J} \rho_i(u_j, v_j)\, \sigma(s_j, t_j)\, \bar{L}(\vec{o}_j, \vec{e}_j)}{\sum_{j<J} \rho_i(u_j, v_j)\, \sigma(s_j, t_j)},
    \label{eq:prefilter-mc}
\end{equation}
where the $(u_j, v_j)$ and $(s_j, t_j)$ are random points on the disk and pixel reconstruction basis function respectively, and the ray $(\vec{o}, \vec{e})$ passes through the 3D location of these two points.

The offline light field ray query, $\bar{L}(\vec{o}, \vec{e})$, used in Equations~\ref{eq:prefilter-integral} and~\ref{eq:prefilter-mc} should be implemented using the highest quality light field reconstruction possible.
In our implementation, we use a reconstruction very similar to the $L(\vec{o}, \vec{e})$ described in Section~\ref{sec:reconstruction}, but we use a higher density set of light field images and the depthmaps before simplification (see Subsection~\ref{sec:mesh-gen}) for the geometry proxy for each light field image.

Figure~\ref{fig:prefilter} shows the rays used to evaluate Equation~\ref{eq:prefilter-mc} for one pixel.  In this example, we use a box filter for $\sigma(s_j, t_j)$, so all rays passing through the pixel are computed.  When the real-time model geometry is close to the shape of the offline model as in Figure~\ref{fig:prefilter}(a), the color samples returned by $\bar{L}(\vec{o}_j, \vec{e}_j)$ will be in close agreement, and the result stored to the prefiltered image $\tilde{I}'_i$ will be sharp.  But when the geometries disagree as in Figure~\ref{fig:prefilter}(b), the rays spread to different parts of the light field images, and the results are blurred.

\section{Light Field Compression Using VP9}
\label{sec:compression}


We generally use $\sim$1,000--1,500 images at 1280$\times$1024 resolution to represent a panoramic light field that can be rendered at high quality using the algorithm of Section~\ref{sec:reconstruction}.  Without compression, this would require 4--6 GBs of image data which would be inconveniently slow to download over a standard 10-20 Mbit/s connection.  Fortunately, the images in a light field exhibit significant coherence which can be leveraged for compression.  

Since our light fields are shot as a sequence of images, one can imagine compressing them with a standard video compression algorithm.  However, typical video compression techniques are problematic for our tile streaming architecture (Section~\ref{sec:tile-streaming}) due to their use of \emph{motion compensated prediction (MCP)}~\cite{sullivan2012hevc,VP9}, where blocks of pixels are predicted from translated pixel blocks earlier in the sequence.  Sadly, this prevents fast random access to arbitrary image tiles within the light field.



Instead, we adapt MCP to support our light field rendering technique (Section~\ref{sec:lf-compress}), and modify the open-source VP9~\cite{VP9} codec to support this (Section~\ref{sec:vp9}).

\subsection{Light Field Compression using MCP}
\label{sec:lf-compress}



\begin{figure}
    \centering
    \includegraphics{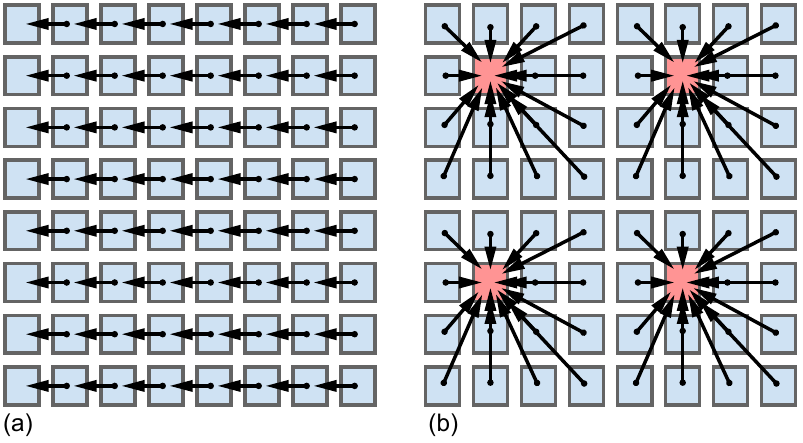}
    \caption{Reference structure used for MCP for light fields.  The blue boxes represent light field images organized in a 2D grid.  (a) A direct application of video MCP to light field data.  Each row is a video GOP where each image uses it's left-hand neighbor as a reference for MCP.  (b) Our approach to MCP for light fields.  We choose a sparse set of images to be references for the entire light field (red boxes).  Every other image uses its nearest reference image for MCP.}
    \label{fig:lightfield-mcp}
\end{figure}

Since we'd still like to use MCP to exploit the coherence between neighboring light field images, we build a reference structure which determines which images are used to predict which other images.



A simple approach (Figure~\ref{fig:lightfield-mcp}(a)) would be to treat whole groups of images as a \emph{group of pictures} (GOP) and encode them as a standard video stream~\cite{dai2015lenselet}.  In Figure~\ref{fig:lightfield-mcp}(a), each row of images is a GOP where each image uses its left-hand neighbor as a reference for MCP.  This leads to long chains of references, which is problematic for our rendering algorithm, since we require fast random access to individual image tiles.  For example, to decode a tile from an image in the right column of Figure~\ref{fig:lightfield-mcp}(a), we would first need to decode every other image to the left, which is inefficient.

Instead, we choose a sparse number of images to be reference images for the whole light field as in Figure~\ref{fig:lightfield-mcp}(b).  Every other image uses the nearest reference image as its reference for MCP.  This guarantees that the reference chains have a length of at most one.  In practice, we decode the reference images when loading the light field from second storage and keep them in memory, allowing us to decode any tile from any other image immediately.


Figure~\ref{fig:lightfield-mcp} shows a common situation where the the light field image positions are in a 2D grid, making it straightforward to choose a subset of the images to be reference images.  Since we also wish to support less regular panoramic light field datasets, we use the following simple hierarchical clustering algorithm (other such algorithms could also likely work).  We first build a bounding volume hierarchy (BVH) over the 3D light field image locations until the leaf nodes contain the target number of images per cluster.  Then we choose the center image as the reference, and all the other images in that leaf node refer to it for MCP.  We found that using 25 images per cluster provides good compression for most light fields, and we use that for all examples in this work.


\subsection{Implementation using VP9}
\label{sec:vp9}
We implemented this compression approach using the open source, royalty free VP9 video codec~\cite{VP9} as implemented in~\cite{vp9code}.  We made two changes to the VP9 code to adapt it for light field compression:

\begin{enumerate}
  \item allow access to more MCP reference images than the 8 provided as default, and
  \item provide random access to individual image tiles.
\end{enumerate}

We accomplished (1) by adding a function to redirect the pointer for one of the internal VP9 reference images to an arbitrary memory buffer.  We call this alternate function to point to any reference image in memory before encoding an image or decoding a tile.  With this change, we can implement any MCP reference structure we choose, including the one shown in Figure~\ref{fig:lightfield-mcp}(b).  

To implement (2), we add a look up table to quickly find the location of a tile in the compressed image data.  The look up table must be compact to avoid introducing too much overhead.  We use a two-level look up table where
the first level points to the start of the column for the tile, and the second points to the tile itself.  This table is added to the image header and entropy encoded,  
adding on average just 1 byte per tile and contributing negligible overhead to decoding.


We must finally ensure that that each tile is encoded completely independent of the others.  This required turning off loop filtering between tiles (although we still use loop filtering to smooth boundaries between blocks within each tile) and turning off the probability adaptation for entropy encoding that depends on neighboring tiles.  




\section{Light Field Processing and Rendering System Details}
\label{sec:rendering}
In this section, we tie together the information in the previous disjoint Sections~\ref{sec:reconstruction}, ~\ref{sec:prefilter}, and ~\ref{sec:compression} to show how it all fits together into a unified system for acquiring, processing, and rendering light fields.  We've already fully described the acquisition stage in Section~\ref{sec:acquisition}, but we haven't yet addressed all of the pieces in processing and rendering.   We fill in the gaps for processing below in Subsection~\ref{sec:processing} and for rendering in Subsection~\ref{sec:subrender}.

\subsection{Processing Details}
\label{sec:processing}
Our full processing pipeline has 8 stages: \emph{calibration} $\rightarrow$ \emph{synthesis} $\rightarrow$ \emph{subsampling} $\rightarrow$
\emph{equiangular conversion} $\rightarrow$ \emph{geometry generation} $\rightarrow$ \emph{prefilter} $\rightarrow$ \emph{color enhancement} $\rightarrow$ \emph{compression}.
Some of the processing stages are covered in previous sections:  geometry generation in Subsection~\ref{sec:mesh-gen}, prefiltering in Section~\ref{sec:prefilter}, and compression in Section~\ref{sec:compression}.  Here we briefly describe the remaining stages.
Note that our pipeline is a fully automated process, as it is intended to scale to many light fields.

\paragraph{Calibration}
The first step of processing is calibration, where we determine the camera extrinsics and intrinsics used for all of the light field images.  This step must be run on all light fields acquired by our camera rigs (see Section~\ref{sec:acquisition}) because camera extrinsics, and even intrinsics, are different for every light field capture.  We use an implementation of the approach described by Wu~\shortcite{wu2013towards} which we've found to be robust as long as we input  approximate priors for the camera extrinsics and accurate priors for the cameras' focal lengths.  As a final part of the calibration step, we warp all images to a rectilinear image space, removing any fisheye or other distortions.

\paragraph{Synthesis}
Our synthesis step creates new light field images to fill the top and bottom poles of the capture sphere because those are blind spots in our camera rigs.  The bottom pole is obstructed by the tripod, and the top is sometimes missing images because, in order to save time, we don't always wind up the rig to point straight up.  We, therefore, use offline view synthesis to interpolate from other images in the light field to cover the poles.  In the light fields shown in this paper, we use the view synthesis algorithm described in Anderson et al.~\shortcite{anderson2016jump}.

\paragraph{Subsampling}
Our light field camera rigs provide overly dense sampling near the top and bottom of the sphere (see Figure~\ref{fig:teaser}(b,d)).  To provide a more regular sampling, we remove some images from these denser regions.  This reduces the size of the light field data, makes rendering faster, and provides more uniform quality throughout the light field.  We use the densest set of light field images as input into the prefilter stage, and output to the subsampled set of images so that we still get benefits from these extra images.

\paragraph{Equiangular conversion}
Our equiangular conversion step converts the input rectilinear images into a locally linear approximation of an equiangular image space.  We use an approximation of one face of an equiangular cube map (EAC)~\cite{eac}.  This allows us to use cameras with a large field of view without needing to pass inordinately large images to the renderer.  
In the examples in this paper, we use 1280$\times$1024 images with 120$^{\circ}\times$96$^{\circ}$ field of view for our DSLR captures and 110$^{\circ}\times$88$^{\circ}$ for our GoPro captures which, using our equiangular approximation, results in a pixel density of $\sim$10--11 pixels per degree which matches the resolution of the current HTC Vive and Oculus Rift displays.  Using regular rectilinear images, it would require 2117$\times$1357 to consistently achieve comparable pixel density.

The equiangular space is a non-linear space which would make it more difficult to render our per-view geometries if directly represented in this space.  Therefore, we only warp the tile corners to their equiangular locations and use standard rectilinear mapping within each tile.  We refer to this as a locally linear approximation of equiangular space, and it gives us almost all of the benefit without incurring significant rendering cost.

\paragraph{Color enhancement} As the final step before compression, the color enhancement stage applies a generic color look-up-table (LUT) to the light field images.  This is usually used to convert from the camera's vendor specific color space into display-ready sRGB.  It can also be used to feed in hand-made LUTs to artistically enhance the colors in the scene.

\paragraph{Cloud processing}
Cloud computing is vital for processing our light field data.  By processing in the cloud in parallel across the images, our pipeline generally takes about 3-4 hours to process a single panoramic light field still.  Running in serial on a single workstation would take over a month.  We have not heavily optimized the pipeline and believe that with further optimizations the total processing time could be reduced significantly.

\subsection{Rendering Details}
\label{sec:subrender}
The architecture of our renderer is driven by the tile streaming and caching algorithm described in Subsection~\ref{sec:tile-streaming}.  On the CPU, we perform tile visibility culling, decode visible tiles that aren't in the cache, and stream the decoded image tiles to the GPU tile cache and the visble triangles to the GPU render pipeline.  Finally, the GPU performs our disk-based reconstruction as described in Subsection~\ref{sec:gpu}.

\paragraph{Parallel decoding}
We run multiple decoder threads on the CPU that decode tiles in parallel.  During normal user head motion, the number of tiles decoded per render frame is quite small, generally less than 50 which is comfortably decodable in the 11ms time budget to maintain 90Hz.  
Most tiles needed to render a new frame are already in the cache.  
However, when loading a new light field, or under especially vigorous head motion, the number of tiles needed can jump to the thousands.

\paragraph{Two-level progressive rendering}
In order to avoid dropping frames due to spikes in tiles to decode, we use the disks from our disk-based representation to perform incremental progressive refinement.  We use a 2-level hierarchy of disks, one coarse and one fine.  Each level is a full and valid light field.  The coarse level is small enough such that all of the light field images in that level can be pre-loaded to GPU RAM and kept there.  During rendering, we loop through all of the visible fine level disks, decode the textures for those images, and send the geometry and texture data to the GPU.  If there isn't enough time in a given frame to decode all of the texture data needed for some of the fine-level disks, then we also draw the coarse level disk in that region.  Progressive rendering has been used to render light fields before~\cite{sloan1997time,birklbauer2013rendering}.  Our disk-based representation makes for an easy implementation as we can render any combination of fine and coarse disks, and they blend naturally in the output.  We generally set the maximum budget of tiles to decode per frame to 128, which is decodable in well under 11ms on most current VR ready CPUs. 

\paragraph{Tile cache}
Our compression algorithm described in Section~\ref{sec:compression} outputs tiles in YUV420 format.  Since this isn't natively supported on all OpenGL platforms, we use two array textures to implement our tile cache, one single component array texture for the y-component and another half-resolution two component array texture for the uv-components.  The y and uv values are sampled separately in the shaders and recombined to output RGB colors.  


In order to either use a smaller memory footprint or be able to store more tiles at the same footprint, our renderer also supports on-demand transcoding of the light field tiles to the RGTC texture format, which reduces the tile cache's GPU memory by 4$\times$ with no visible loss in quality.  This does add about 30\% performance overhead to every tile decode.  We generally limit the tile cache size to 2GB.

\paragraph{Valid view volume fade}  Our panoramic light field datasets provide the full set of rays needed to generate novel views from within a limited viewing volume.  Outside of that viewing volume, the edges of the light field images become visible which leads to distracting artifacts.  Rather than show this directly to the user, we smoothly fade out the light field as the user approaches the edge of the viewing volume.  
\section{Results}
In this section, we evaluate our system in terms of rendering quality, file sizes, and rendering speed\rev{; the technical challenges (3)--(5) that were introduced in Section~\ref{sec:intro}}.

\paragraph{Test Scenes}

Table~\ref{tab:scenes} presents acquisition statistics for our test scenes.  All scenes, except for \emph{Lab Space}, were acquired using the camera rigs described in Section~\ref{sec:acquisition} and have representative rendered views in Figure~\ref{fig:teaser}.   We created \emph{Lab Space} specifically for Figure~\ref{fig:plane-place} using a robot arm programmed to move a DSLR camera and take pictures at points on the sphere.  This allowed us to acquire the light field using an arbitrarily dense sampling of the sphere.

\label{sec:results}
\begin{table}[ht]
    \centering
    \caption{Statistics for the test scenes used in this paper.}
    \includegraphics{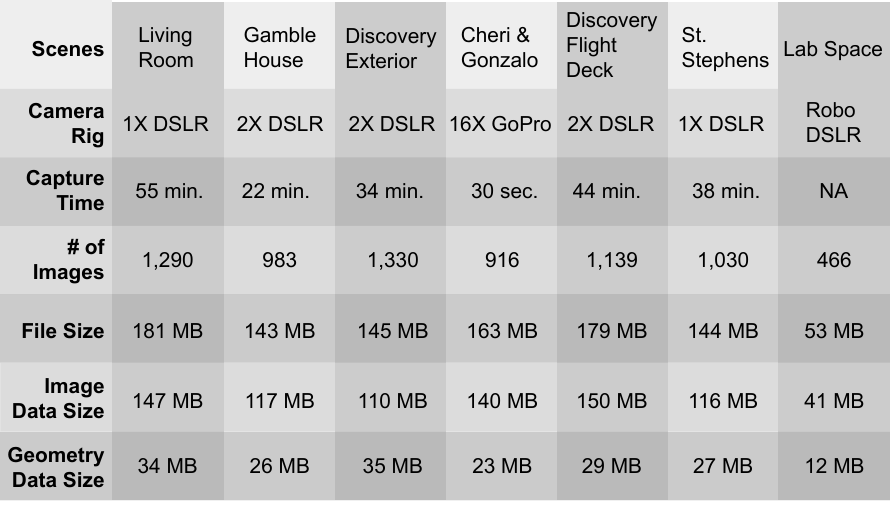}
    \label{tab:scenes}
\end{table}

\subsection{Render Quality}

\subsubsection{Comparison to Other Real-time Reconstruction Algorithms}
We compare our rendering algorithm to two other real-time light field rendering algorithms:  camera mesh blending field with planar geometry proxy (\emph{CMBF+P})~\cite{levoy1996light} and camera mesh blending field with a global geometry proxy (\emph{CMBF+GG})~\cite{davis2012unstructured}.
\begin{figure*}
    \centering
    \includegraphics{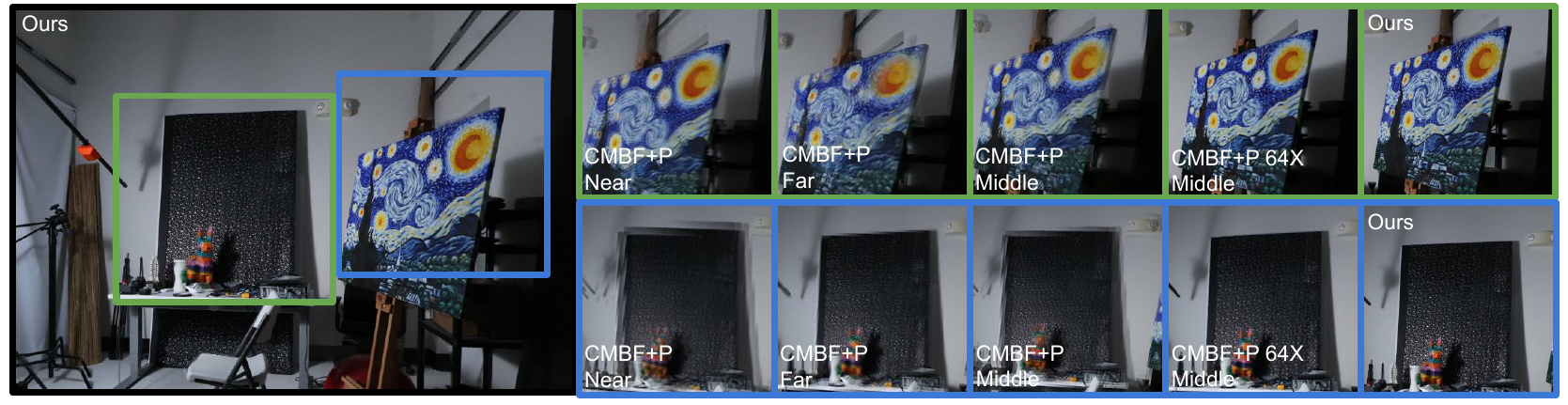}
    \caption{Our renderer compared to camera mesh blending field with planar geometry proxy (\emph{CMBF+P}).  On this scene, our approach achieves sharp, high quality results with only 466 light field images.  With CMBF+P, there are visible ghosting artifacts whether we place the plane to align with \emph{Near}, \emph{Far}, or \emph{Middle} scene geometry.   CMBF+P requires at least 29,722, or $\sim$64$\times$ more,  light field images to achieve comparable results (CMBF+P 64$\times$).}
    \label{fig:plane-place}
\end{figure*}

\begin{figure*}
    \centering
    \includegraphics{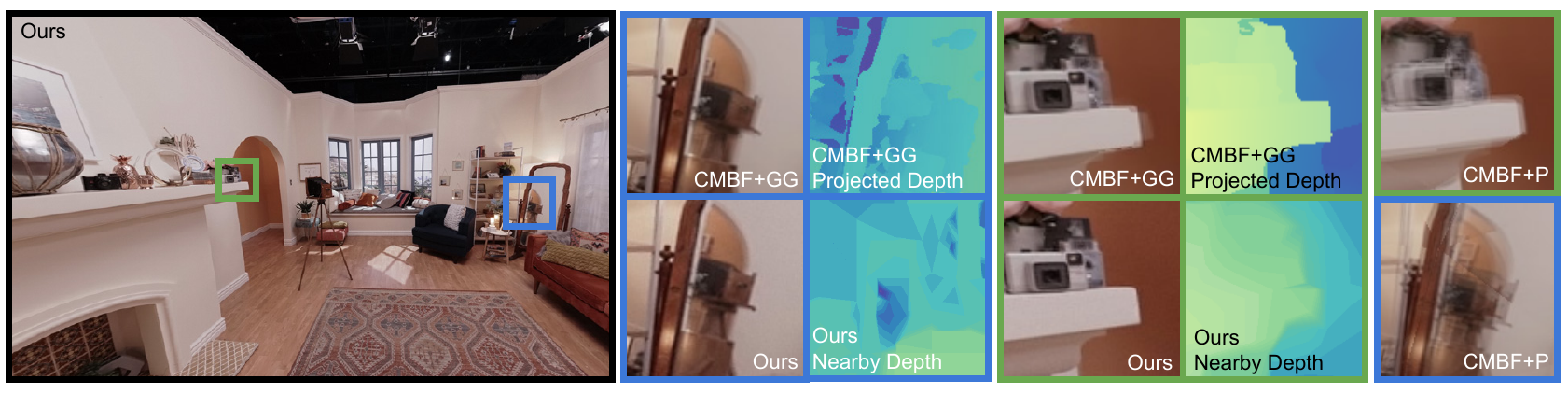}
    \caption{Our renderer compared to using a camera mesh blending field with a global geometry proxy (\emph{CMBF+GG})~\cite{davis2012unstructured}.  Our approach achieves high quality results with relatively coarse per-view geometry.  Our rendered results are shown along with the simplified geometry from a light field image near the rendered view.  CMBF+GG rendered results are shown along with the global geometry projected onto the rendered view.  CMBF+P is shown at the far right, where the plane is aligned with the middle of the scene.}
    \label{fig:vs-gg}
\end{figure*}
\paragraph{Camera Mesh Blending Field with Planar Proxy (CMBF+P)}
Figure~\ref{fig:plane-place} compares our technique to CMBF+P as in Section~\ref{sec:vs-camera-mesh}.  Although CMBF+P is not state-of-the-art, it provides a useful point of comparison.  Our method achieves high quality reconstruction with only 466 light field images.  For CMBF+P, the placement of the plane greatly affects the reconstruction quality.  Unfortunately, with only a single plane and 466 light field images, there is significant ghosting regardless of placement.  

In general, a more accurate geometry proxy means fewer images required to render the light field without artifacts~\cite{lin2004geometric}.  In order for CMBF+P to approach our reconstruction quality, we had to create a light field with 29,722 images, $\sim$64$\times$ more than our method.

\paragraph{Camera Mesh Blending Field with Global Geometry (CMBF+GG)}
We also compare to CMBF+GG, which is similar to CMBF+P except it uses a global geometry model instead of a plane.  Our CMBF+GG renderer is based on the algorithm introduced by Davis et al.~\shortcite{davis2012unstructured}.  Our implementation differs in that we focus on inside-out panoramic light fields instead of outside-in; Davis et al. also provides an interactive approach to generate the global geometry.

By far the largest benefit of our per-view geometry approach is that it is significantly easier to generate per-view depth maps which are good enough to reconstruct local views than it is to create global geometry accurate enough to use for all views.  We found it difficult to build a geometric model of the scene using photogrammetry software (RealityCapture~\cite{realitycapture}) directly from our outward-looking light field imagery.  We therefore performed a more traditional photogrammetric acquisition of the space by moving a hand-held DSLR camera with a rectilinear lens throughout the space.  This produced a reasonable model, but required an alignment process to register it to our light field.

Figure~\ref{fig:vs-gg} compares the results using our method and CMBF+GG.  CMBF+GG is a significant improvement over CMBF+P.  However, our CMBF+GG implementation struggles near edges because it is hard to create a model that aligns closely to the edges for \emph{all} views.  In order for CMBF+GG to achieve high reconstruction quality, it would require near perfect calibration, geometry generation, and registration which is difficult to achieve in practice.  Our approach is more tolerant of errors in calibration and geometry generation as can be seen with the depth image comparison in Figure~\ref{fig:vs-gg}.

Reflective and other non-diffuse surfaces are also difficult for CMBF+GG because photogrammetry cannot construct a model that is consistent for all views.  The per-view geometries used by our method only need to be a reasonable approximation for nearby views and thus can handle reflective surfaces reasonably well.



\subsubsection{Prefiltering}
\begin{figure*}
    \centering
    \includegraphics{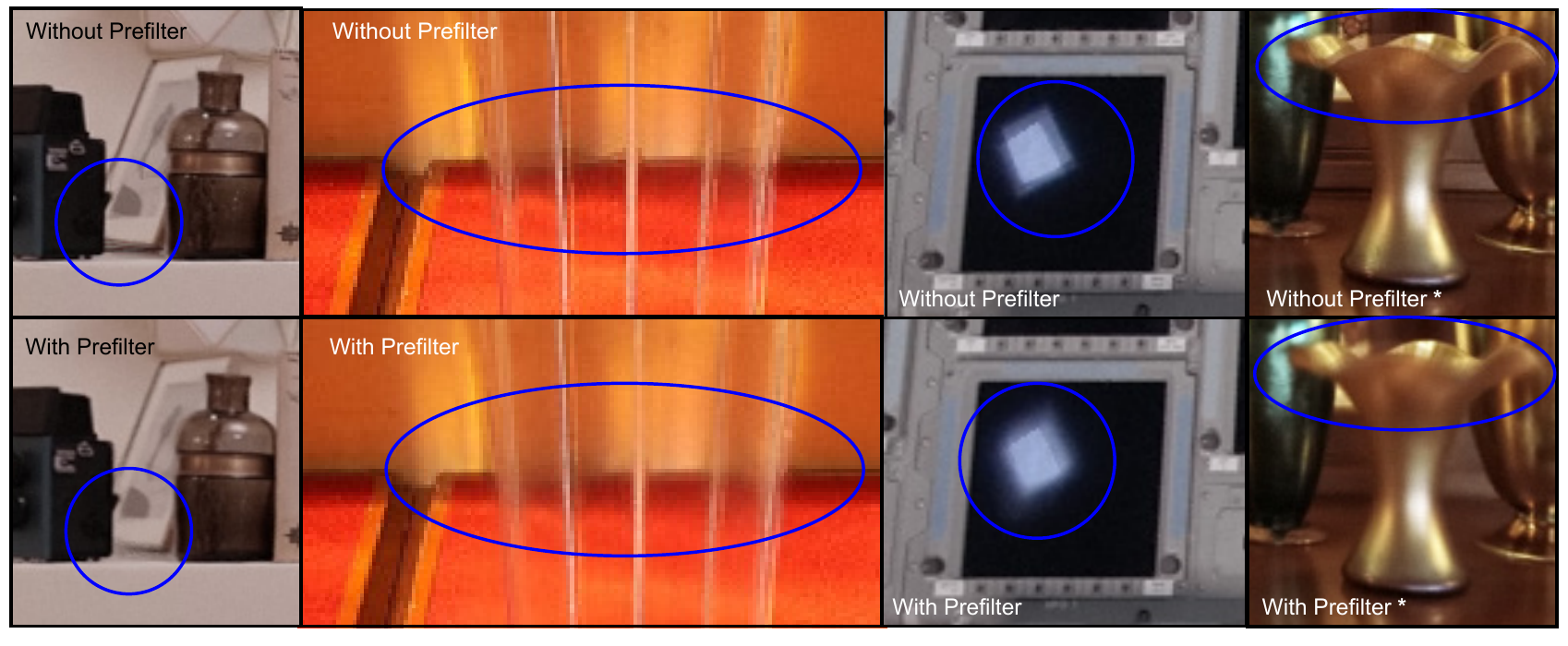}
    \caption{Prefiltering blurs over distracting ghosting artifacts which occur where the geometry proxy is inaccurate.  The scenes used are (left to right) \emph{Living Room}, \emph{St. Stephens}, \emph{Discovery Exterior}, and \emph{Gamble House}.  * We use the lower quality geometry setting of gp=1 for this \emph{Gamble House} example to make the prefilter results more visible.  We otherwise use gp=5 to maximize quality.}
    \label{fig:prefilter-results}
\end{figure*}
Our prefiltering algorithm, described in Section~\ref{sec:prefilter}, mitigates the visible impact of light field rendering artifacts.  The primary artifact typically seen in light fields is ghosting, which is particularly distracting in a real-time setting because it appears to flicker as the user moves.  Our prefiltering step replaces many of these ghosting artifacts with a smooth blur which reduces flickering.

Figure~\ref{fig:prefilter-results} shows close-ups of some regions most impacted by prefiltering.  Prefiltering preserves sharp imagery where the per-view geometry is accurate to the scene; elsewhere, it smooths away distracting ghosting artifacts.  

The visual effect of prefiltering can appear subtle when viewed in still images, especially when the per-view geometry is accurate which it tends to be.  Therefore, we used a lower quality geometry for the far right image in Figure~\ref{fig:prefilter-results} to make the difference more clear.  The benefits of prefiltering can also be seen in the accompanying video.

\subsection{Size: Compression Evaluation}
Here, we evaluate the results of the compression algorithm described in Section~\ref{sec:compression}.  We use PSNR to measure compressed image quality. 
Like many image and video codecs, VP9 starts by converting from RGB to chroma-subsampled YUV420 space.  This chroma subsampling is considered visually lossless, and so PSNR is measured in YUV420 space as follows:
$$SSE_{C,s}=\sum_{\frac{W}{s},\frac{H}{s}}[C_1(m,n)-C_2(m,n)]^2,$$
$$PSNR = 10 \log_{10}\bigg( \frac{255^2\cdot \frac{3}{2}W H}{SSE_{Y,1}+SSE_{U,2}+SSE_{V,2}} \bigg) ,$$
where $Y$,$U$,and $V$ are the corresponding planes of each $W$$\times$$H$ image.

\begin{figure}[b]
    \centering
    \includegraphics[width=0.45\textwidth]{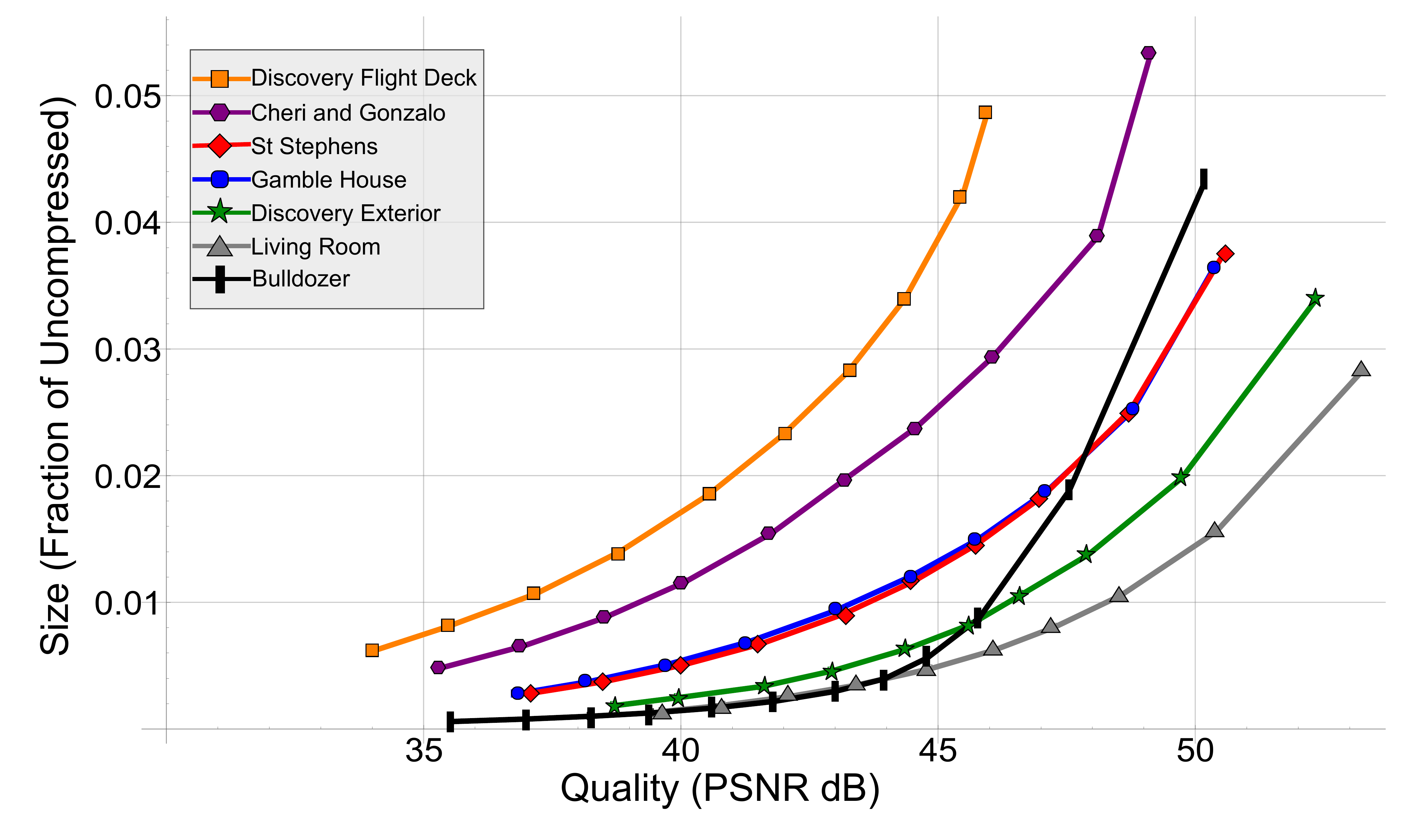}
    \caption{Rate-distortion plot of our compression approach on our test scenes.  We measure size as a fraction of the uncompressed dataset size, which is 24 bits per pixel RGB.}
    \label{fig:rd}
\end{figure}

We refer to Figure~\ref{fig:rd} to analyze the rate-distortion characteristics of our codec.  This plot includes our 6 scenes in Figure~\ref{fig:teaser} as well as the \emph{Bulldozer} from the Stanford Light Field Archive~\cite{vaish2008new}.  Considering only our own scenes (we discuss \emph{Bulldozer} below) at very high quality ($\sim$45 dB PSNR), most light fields are compressed by $\sim$40$\times$--200$\times$.  The one exception is the \emph{Discovery Flight Deck} at $\sim$29$\times$, which is a particularly difficult scene because it has many sharp details on the consoles which are close to the camera.  At more moderate quality (38--40 dB PSNR), we achieve compression rates up to 735$\times$.  

\begin{figure}
    \centering
    \includegraphics[width=0.45\textwidth]{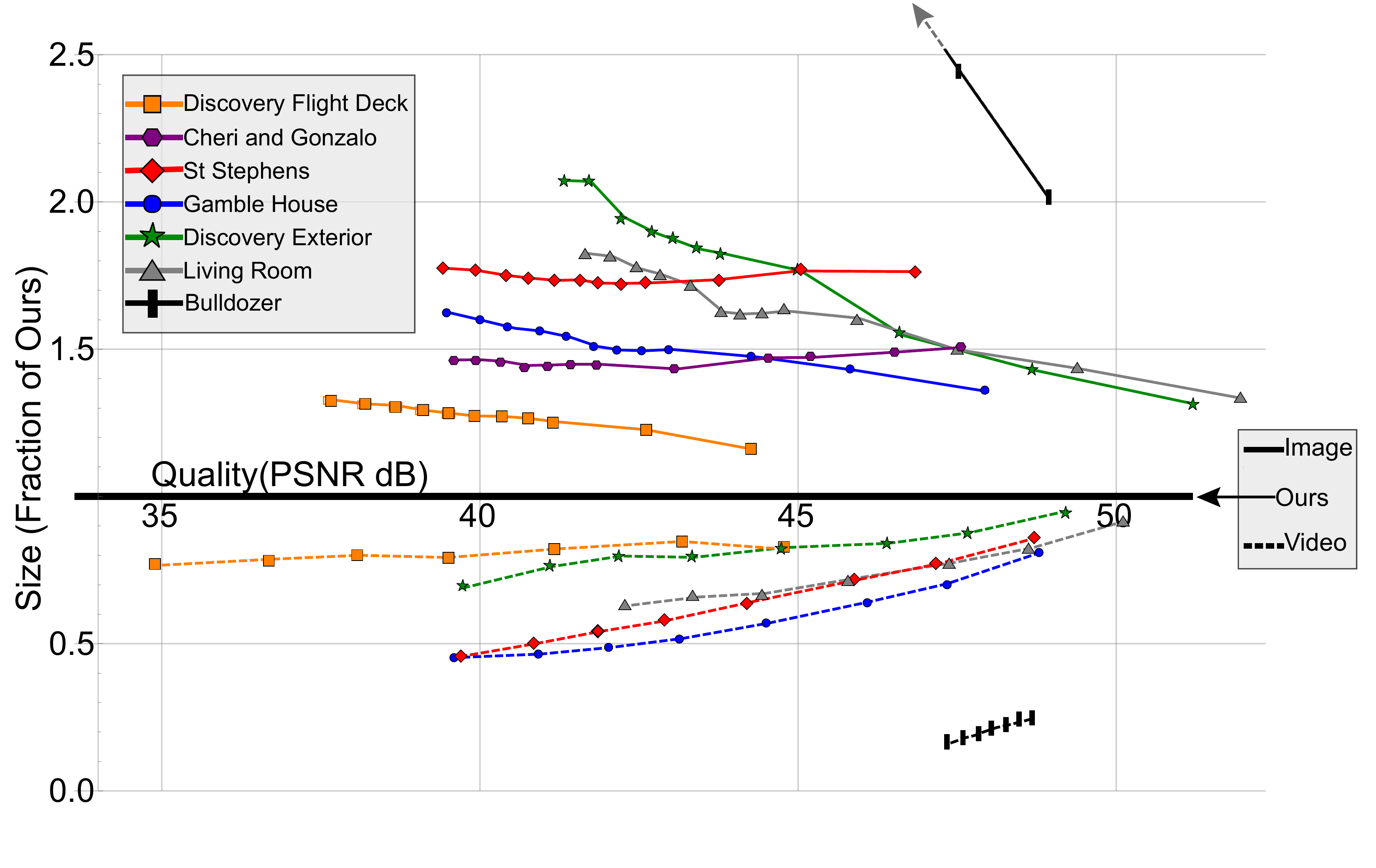}
    \caption{Our compression vs. standard image (solid lines) and video compression (dashed lines).  File sizes are expressed as the scale relative to our file size at the same PSNR.  Thus our size is the 1.0 line.  Note that the line for \emph{Bulldozer} with image compression grows too large for the range at 47 dB and hits 7$\times$ at 40 dB PSNR outside of the displayed graph.}
    \label{fig:vsvp9}
\end{figure}

Figure~\ref{fig:vsvp9} shows the effect of MCP, comparing our results against standard image and video compression, using WebP~\cite{webp} to compress the images and VP9 to compress the videos.  Comparing to image compression shows how much we benefit from MCP, and comparing to video compression shows how much we lose by adapting MCP to support random access to tiles. To arrange our light field images into linear videos, we follow the order in which they were acquired by each camera.

Our compression performance lies somewhere between image and video compression.  Relative to image compression, both ours and the video codec increasingly benefit from MCP with lower quality settings, whereas on our scenes we achieve $\sim$1.5$\times$--2$\times$ better compression than standard image compression, except for the difficult \emph{Discovery Flight Deck}.  Compared to video, we see that there may be more to be gained from a denser reference structure, but for our scenes not more than $\sim$2.1$\times$.  

\emph{Bulldozer} is an especially dense planar light field, and accordingly yields greater compression improvement from MCP.  The curve for \emph{Bulldozer} in Figure~\ref{fig:rd} is much steeper: at 45 dB PSNR, we achieve 178$\times$ compression, and at 38 dB, it is 1,000$\times$.  Moreover, compared to image compression in Figure~\ref{fig:vsvp9}, the \emph{Bulldozer} curve grows too large for the graph range at $\sim$47 dB and hits 7$\times$ at 40 dB.  Even then, the curve for video compression shows that our sparse MCP scheme could benefit from a denser reference strategy.  

\rev{The \emph{Bulldozer} results also illuminate some competition between our rendering and compression algorithms.}  As noted above, our disk-based reconstruction with per-view geometry allows us to render high quality light fields with more sparsely spaced images than \emph{Bulldozer}.  This itself is a valuable form of compression, but it reduces the coherence that is available to our compression algorithm and MCP in general.

\subsection{Speed}
\begin{table}
    \centering
    \caption{Performance statistics for our light field renderer.  All values are averages per frame recorded over a 30 second scripted session.}
    \label{tab:performance}
    \includegraphics{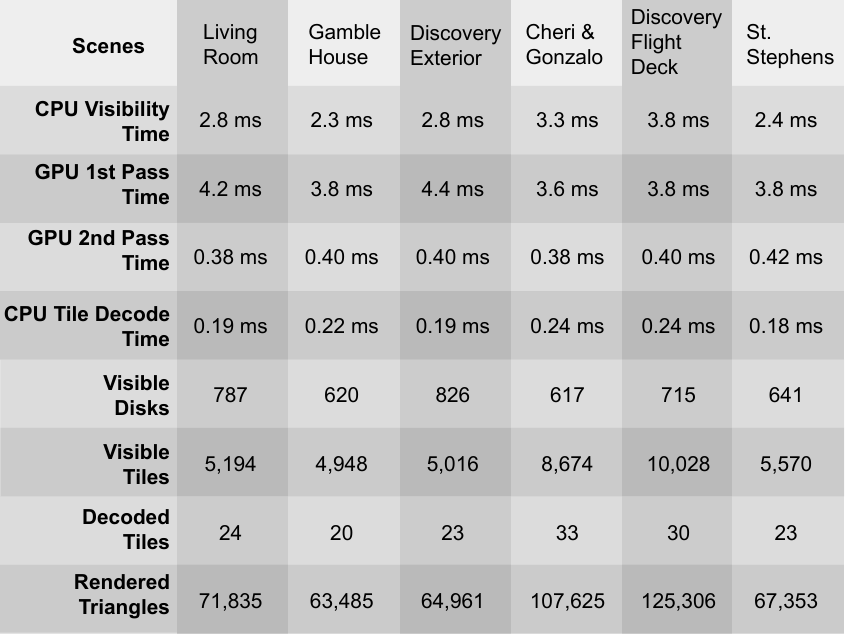}
\end{table}
We study the speed of our light field renderer using Table~\ref{tab:performance}.  To collect these statistics we ran a 30 second scripted session \rev{that mimics a very active user with a wide variety of head motion}.  We ran the sessions on a 2017 Razer Blade VR laptop with an Oculus Rift HMD.  This system has an Intel Core i7-7700HQ CPU running at 2.8 GHz with 16 GB system RAM and an NVIDIA GeForce GTX 1060 GPU with 6GB RAM.  The Oculus Rift renders to two 1080$\times$1200 OLEDs, one per eye, for a total of 2160$\times$1200 pixels.

As seen in Table~\ref{tab:performance}, average render times fit comfortably within the 11 ms per frame budget to maintain 90 Hz.  \rev{This table shows the time spent in each of the multiple phases of our rendering algorithm.  Overall performance is most directly tied to the number of \emph{Visible Tiles} per frame.}  On scenes where the geometry is relatively close to the viewer, such as \emph{Cheri \& Gonzalo} and \emph{Discovery Flight Deck}, more tiles are rendered which leads to increased \emph{CPU Visibility Time}.  \emph{CPU Visibility Time} includes all time culling tiles and aggregating the visible tiles' triangle meshes (see Subsection~\ref{sec:tile-streaming}).  Only 20--33 tiles are decoded per frame on average, even though $\sim$5,000--10,000 are rendered, demonstrating the effectiveness of our tile cache.  The cache is helped by the fact that a person generally can't move their head very far in only 11 ms.

\subsection{Limitations}
Using our system, we are able to acquire, process, and render high quality light fields, but limitations remain.  The most common suggestion from users is that the size of the viewing volume feels somewhat constraining.  To address this, the latest version of our 2xDSLR rig has a $60\%$ longer platform and acquires light field ray volumes over 1 meter in diameter.

Our light field renderer produces high-quality results, but as can be seen in Figure~\ref{fig:prefilter-results}, some subtle artifacts remain.  \rev{As predicted by the analysis in Chai et al.\shortcite{chai2000plenoptic} and Lin and Shum\shortcite{lin2004geometric}, artifacts  appear when our geometry doesn't match the scene, especially when the scene geometry is close to the camera.}  The most noticeable artifacts are around edges at large depth discontinuities.  Our depth-based geometry representation creates a rubber sheet effect at these boundaries.  Our second biggest problem is with surfaces that aren't well modeled by a single geometry layer, such as surfaces with glossy reflections on textured surfaces.  Our geometry model must choose one layer to model well, while the other layer will exhibit some degree of ghosting.  Our prefiltering step helps make these artifacts less distracting, replacing one artifact with a relatively unobjectionable blurring effect.  An explicit multi-layer model as in \rev{\cite{zitnick2004high,penner2017}} could improve rendering quality at edges, though there would be some performance cost to rendering multiple layers per image.  \rev{We could also collect denser light fields and rely less on geometry and view interpolation, but this would lead to increased burden on both the camera rig and the compression algorithm.}

\rev{Our system is intended to scale to all scene types regardless of complexity.  However, our moving light field camera rig limits us to static scenes.  We are able to capture people as long as they remain still for 20 seconds, but movement such as trees swaying in wind produce noticeable ghosting artifacts.}

\rev{There is a lot of room for optimization in our offline processing pipeline.  We achieve reasonable run times (3--4 hours) by relying on massive parallelization using Google's extensive cloud resources, but the run time would be prohibitive for those who don't have access to such resources.}

\section{Conclusion and Future Work}
\label{sec:conclusion}

We have described a complete system for acquiring, processing, and rendering light field panoramic stills for VR.  Our novel camera hardware is versatile, capable of acquiring high quality light fields under a variety of conditions.  It is also portable and time-efficient.  
Our novel light field reconstruction algorithm generates higher quality views with fewer images than previous real-time approaches, and we further improve reconstruction quality using a novel application of light field prefiltering.  Our new light field codec compresses high quality light field stills down to reasonable sizes, and our renderer decodes the images on-demand and reconstructs stereo views consistently at 90 Hz on commodity hardware, a requirement for comfortable VR.

Taken together, we have an end-to-end solution for distributing light field experiences to end users.  We have launched this system to the public \rev{as \emph{Welcome to Light Fields}, a freely downloadable application on the Steam$\circledR$ store (https://store.steampowered.com/)}.  The application includes our light field renderer and library of light field panoramic still images, and it has been downloaded over 15,000 times.


Looking beyond light field stills, the most obvious direction to take our work is into light field video.  Building a practical light field video camera rig remains a challenge.  Such a rig could require as many as a hundred cameras, which is difficult to achieve if we also require it to be portable and easy to use. 

Compression is also clearly more challenging for video. We believe our light field stills codec provides a useful foundation on which to build toward video. Our biggest challenge is decoding the light field video frames fast enough to maintain 90 Hz rendering speed.  This might be easiest to achieve through a hardware implementation of the decoder, and reducing the number of tiles required to render a given frame.

\begin{acks}
This paper and the system it describes are the product of a multi-year effort with support from many people at Google.  In particular, we'd like to thank Jason Dourgarian for help generating figures for the paper; Matthew DuVall for creating the video; Jiening Zhan, Harrison McKenzie Chapter, Manfred Ernst, Stuart Abercrombie, Kay Zhu, Sai Deng, Pavel Krajcevski, Yun Teng, and Graham Fyffe for contributions to the light field processing and rendering software; Xueming Yu and Jay Busch for building the light field camera rig hardware; the WebM team, especially Yunqing Wang, Jim Bankoski, Yaowu Xu, Adrian Grange, and Debargha Mukherjee, for help designing and implementing light field compression; Katherine Knox for project management; Marc Levoy, Janne Kontkanen, Florian Kainz, John Flynn, Jon Barron, Eric Penner, Steve Seitz, and Hugues Hoppe for helpful conversations, guidance, and feedback; Clay Bavor for his consistent support and vision for the project; and Greg Downing who managed the acquisition for most of the light fields used in this paper.
\end{acks}

\bibliographystyle{ACM-Reference-Format.bst}
\bibliography{paper} 

\end{document}